\documentclass[10pt,aps,prb,superscriptaddress,
twocolumn,showpacs,floatfix,notitlepage]{revtex4-2}
\usepackage{amsmath}
\usepackage{amsfonts}
\usepackage{amssymb}
\usepackage{hyperref}
\usepackage{graphicx}
\usepackage{color}
\usepackage[mathscr]{euscript}
\usepackage{bbm}
\usepackage{braket}
\usepackage{amsmath}
\usepackage{amssymb}
\usepackage{wrapfig}
\usepackage{tabularx}
\usepackage[utf8]{inputenc}
\usepackage{booktabs}
\usepackage[table]{xcolor} 
\usepackage{siunitx}

\usepackage{natbib}
\usepackage{rotating}
\usepackage[utf8]{inputenc}
\usepackage[T1]{fontenc}
\usepackage{soul}

\hypersetup{colorlinks=true,linkcolor=magenta
  ,citecolor=magenta, urlcolor=magenta}

\begin{document} 
\flushbottom 
\title{Estimation of the second-order coherence function using quantum reservoir and ensemble methods}

\author{Dogyun Ko}
\affiliation{%
	Institute of Physics, Polish Academy of Sciences, Al. Lotnik\'{o}w 32/46, 02-668 Warsaw, Poland}%
\author{Stanisław Świerczewski}
\affiliation{Institute of Experimental Physics, Faculty of Physics,\\University of Warsaw, ul. Pasteura 5, PL-02-093 Warsaw, Poland}
\author{Andrzej Opala}
\affiliation{Institute of Experimental Physics, Faculty of Physics,\\University of Warsaw, ul. Pasteura 5, PL-02-093 Warsaw, Poland}
\affiliation{%
	Institute of Physics, Polish Academy of Sciences, Al. Lotnik\'{o}w 32/46, 02-668 Warsaw, Poland}%
\author{Michał Matuszewski}
\affiliation{%
	Institute of Physics, Polish Academy of Sciences, Al. Lotnik\'{o}w 32/46, 02-668 Warsaw, Poland}%
\affiliation{%
	Center for Theoretical Physics, Polish Academy of Sciences, Al. Lotnik\'{o}w 32/46, 02-668 Warsaw, Poland}%

\author{Amir Rahmani}
\affiliation{%
	Institute of Physics, Polish Academy of Sciences, Al. Lotnik\'{o}w 32/46, 02-668 Warsaw, Poland}%

\begin{abstract} 
We propose a machine learning-based approach enhanced by quantum reservoir computing (QRC) to estimate the zero-time second-order correlation function $g^{(2)}(0)$.  
Typically, measuring $g^{(2)}(0)$ requires single-photon detectors and time-correlated measurements.
Machine learning may offer practical solutions by training a model to estimate $g^{(2)}(0)$ solely from average intensity measurements  
In our method, emission from a given quantum source is first processed in QRC. During the inference phase, only intensity measurements are used, which are then passed to a software-based decision tree-based ensemble model. We evaluate this hybrid quantum–classical approach across a variety of quantum optical systems and demonstrate that it provides accurate estimates of $g^{(2)}(0)$. We further extend our analysis to assess the ability of a trained model to generalize beyond its training distribution, both to the same system under different physical parameters and to fundamentally different quantum sources. While the model may yield reliable estimates within specific regimes, its performance across distinct systems is generally limited.  
\end{abstract}

\date{\today}

\maketitle
\section{Introduction}\label{sec:intro}
\begin{figure*}[t]
    \centering
    \includegraphics[width=0.31\linewidth]{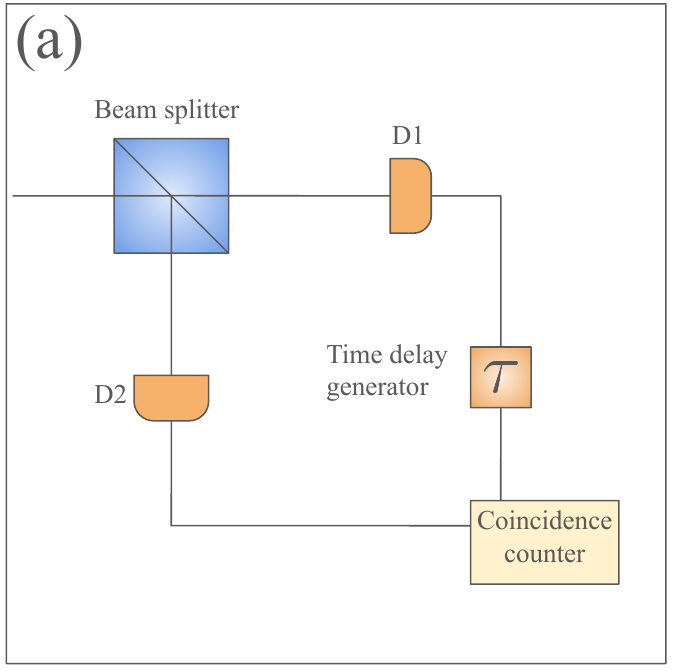}
    \includegraphics[width=0.64\linewidth]{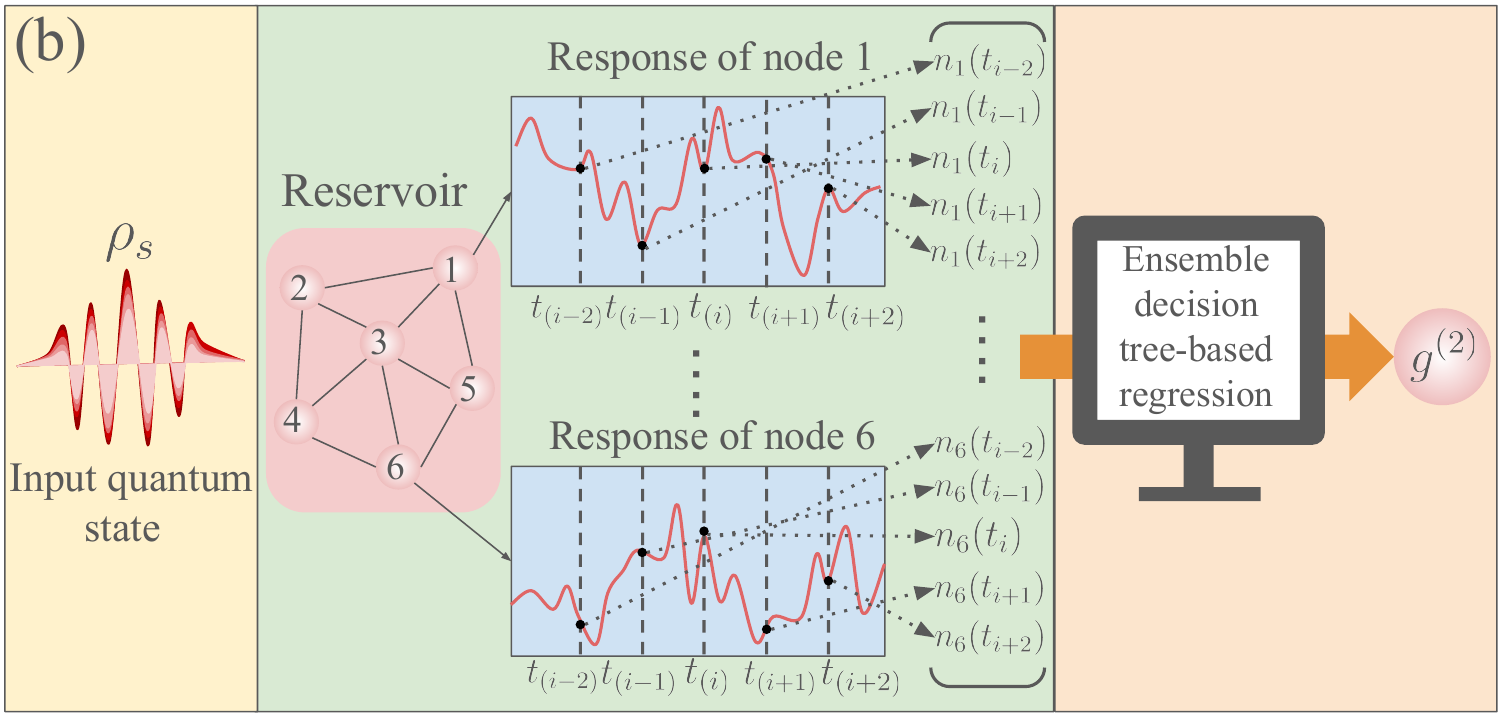}
    \caption{(a) Setup for the Hanbury Brown and Twiss experiment to measure second-order coherence. A signal is sent through a beam splitter and the outputs are incident on two detectors (D$_1$ and D$_2$). One of the beams may experience a relative time delay $\tau$ with respect to the other. The outputs from the detectors are analyzed by a coincidence counter, which generates a pulse whose height is proportional to the time delay $\tau$. (b) Schematic of our proposal to predict $g^{(2)}(0)$ based on enhanced QRC. The quantum optical state (source) is fed into the quantum reservoir. 
    The nodes in the reservoir have their own dynamics, and when they interact with the excitations from the source, they may exhibit oscillations. The average occupation numbers of the nodes are recorded. The last trainable layer is implemented in software, resulting in predictions of $g^{(2)}(0)$. 
    }
    \label{fig:ourproposal}
\end{figure*}

Light emitted by different sources can be characterized by properties such as intensity, spectrum, and polarization. However, a fundamental question may arise: Are there additional ways to distinguish light from different sources beyond these conventional parameters? One approach is to examine the statistical properties of photon emission over time, particularly the temporal correlations in their arrival times. These correlations, which reveal collective photon behavior, are mathematically described by correlation functions in quantum optics—a concept that was experimentally validated in the pioneering Hanbury Brown and Twiss experiment of the 1960s \cite{Brown01071954,Dravnis}. Considering correlations between pairs of photons, the central concept is the (normalized) second-order correlation function defined as \cite{Gerry_Knight_2023}
\begin{align}
    g^{(2)}(\mathbf{r},\tau)=\frac{\langle \hat E^\dagger(\mathbf{r},t)\hat E^\dagger(\mathbf{r},t+\tau) \hat E(\mathbf{r},t+\tau) \hat E(t)\rangle}{\langle \hat E^\ast(\mathbf{r},t) \hat E(\mathbf{r},t)\rangle}\,,
\end{align}
%
with electric field operator $\hat E$ at space-time ($\mathbf{r},t$).  $\tau$ is the time delay and $\langle\cdots\rangle$ means the ensemble average of all possible field realizations (see panel (a) in Fig.~\ref{fig:ourproposal}). To identify the nature of the emission properties of light sources, the zero-delay $g^{(2)}(\tau=0)$ is commonly
 used. It is known that for a thermal source $g^{(2)}(0)=2$, while for a coherent source $g^{(2)}(0)=1$. 
Moreover, the antibunching condition $g^{(2)}(0)<1$ is restricted to nonclassical states of light~\cite{Gerry_Knight_2023}. In general,  $g^{(2)}(0)$ provides information about the photon statistics, in particular quantifying photon bunching and antibunching.

Measuring $g^{(2)}(0)$ can present several challenges, especially for quantum light sources with low intensity, where it may take a considerable amount of time to gather enough photons for accurate statistical measurements \cite{Sempere-Llagostera:22,PhysRevLett109183601}. 
Typically, measuring $g^{(2)}(0)$ requires single-photon detectors and time-correlated measurements. Additional difficulties arise from low collection efficiency due to insufficient timing resolution or the presence of background noise \cite{Krapick_2013}. That said, one may ask whether machine learning can assist in overcoming or alleviating such challenges, allowing for estimation of $g^{(2)}$ or other correlation functions from averaged intensity measurements, allowing for estimation of $g^{(2)}$ or other correlation functions from averaged intensity measurements. 
For instance, if measured data of $g^{(2)}$ is obtained for a given physical system or set of parameters, how can this information be used first to estimate $g^{(2)}$ and then generalize to unmeasured values of $g^{(2)}$ for the same system under different parameters—or even for a fundamentally different physical system? Here, we propose a machine learning framework, enhanced by quantum reservoir computing (QRC) \cite{Ghosh_QRCReview,ghosh2019quantum}, to estimate 
$g^{(2)}(0)$ in the inference phase using intensity measurements only. Quantum reservoir computing is the extension of classical reservoir computing, a peculiar type of neural network \cite{TANAKA2019100}, to quantum systems. Quantum reservoir computing network includes a network of interconnected quantum systems, such as qubits or quantum harmonic oscillators, that can transform input data to higher dimensions of the quantum system Hilbert space. Different kinds of quantum and classical tasks have been explored using QRC~\cite{
Ghosh_QRCReview, fujii2017harnessing,ghosh2019quantum,
mujal2021opportunities,
martinez2021dynamical,
PhysRevResearch.6.013051,
https://doi.org/10.1002/pssr.202400433}.
Our proposal (depicted schematically in Fig.~\ref{fig:ourproposal}) is based on QRC enhanced with a classical boot‐strap aggregating (bagging) algorithm. In particular, we use Random Forest and Extra-Trees, which are an ensemble of Decision Trees, 
trained via the bagging method \cite{randomF,ExTrees}. These algorithms possess significant adaptive learning capabilities and the ability to model nonlinear mappings, enabling their application across a wide range of domains \cite{statistical_learning}. Quantum input states are coupled to the nodes of the reservoir, whose dynamics depend nonlinearly on the input state. In the training phase, emission intensity from the nodes averaged over many realizations is recorded as a function of time,  and then analyzed in software. During training, we assume that $g^{(2)}(0)$ of input states is known from independent measurements. The training consists of learning the parameters of the classical bagging algorithm, while the reservoir is not changed. In the inference phase, intensity measurements of the reservoir nodes allow to estimate $g^{(2)}(0)$ without any physical correlation measurements. In our case, we consider various quantum optical systems as sources, including statistical mixing of continuous variable (CV) states which shows the photon statistics ranging from antibunching to bunching \cite{ NIETO1997135,RevModPhys77513}, squeezed photon-added states \cite{PhysRevLett.98.153603,PhysRevLett.98.153603} which is an example of non-Gaussian state with the different regime of photon statistics upon changing the squeezing,  emitter in the cavity \cite{Delteil2019}, and mixing emission in a beam splitter \cite{math10244794}, as a few examples of a broad class of quantum optical sources.

We show that even with a small reservoir (two nodes), our learning scheme can precisely estimate $g^{(2)}(0)$  without modification of the reservoir, whose parameters remain fixed during training. Interestingly, in some cases, the average occupation number of the source contains enough information to yield accurate estimates of $g^{(2)}(0)$, eliminating the need for full quantum state processing. As we discuss in detail later, this may occur when the occupation number and coherence exhibit similar trends with respect to a given parameter. However, in more demanding cases, a quantum reservoir that is coupled to the quantum states of the source is often required. After training and testing the model on a specific quantum source, we extend our analysis to evaluate its ability to predict $g^{(2)}(0)$ for the same system under varying parameters, as well as for fundamentally different quantum sources. Our results indicate that while the model yields accurate estimates within certain parameter regimes, its generalization ability across distinct systems is generally limited.
Considering the challenge of single-photon detection used to measure $g^{(2)}(0)$, our approach offers an alternative solution for estimating $g^{(2)}(0)$ using enhanced quantum reservoir computing with a nonlinear bagging algorithm. This may provide a new tool for the estimation of coherence functions in quantum optical systems.

The paper is organized as follows. In Section \ref{sec:QRC}, we introduce the general formalism of quantum reservoir computing. Section \ref{sec:res} explores various physical optical systems, detailing the data preparation process for training and presenting the results of estimations. Finally, in Section \ref{sec:disc}, we discuss the generalization properties and the potential for using a trained model to make predictions in both similar and dissimilar physical systems.  

\section{Quantum Reservoir 
computing and classical modeling}\label{sec:QRC}
Reservoir computing is a supervised learning approach based on recurrent neural networks~\cite{SHERSTINSKY2020132306} 
implemented in either physical or abstract dynamical systems \cite{TANAKA2019100}. The word \emph{reservoir} implies a physical system that is a fixed, often randomly connected network of nodes with an internal memory of the past history \cite{jaeger:techreport2001}.  Input data are fed and processed recurrently in the internal states of the reservoir system. Considering quantum states as input, the QRC enables its processing and embeds it into a higher-dimensional quantum space, allowing the extraction of more complex patterns than its classical counterparts. \cite{Hulser:22}. 

The scheme of our proposal is shown in Fig.~\ref{fig:ourproposal}b. The quantum source (such as an emitter embedded in a cavity) is characterized by the initial density matrix $\rho_s$. Each initial state is a sample in our dataset, and we focus on its second coherence function $g^{(2)}(\tau=0)=\frac{\mathrm{Tr}[\hat{a}_s^\dagger \hat{a}_s^\dagger \hat{a}_s\hat{a}_s\rho_s]}{n_{a_s}^2}$, defined with the source mode operator $\hat{a}_s$ and occupation number $n_{a_s}=\mathrm{Tr}[\hat{a}_s^\dagger \hat{a}_s\rho_s]$. The input is coupled to the reservoir through a random coupling constant $\textbf{W}^{\text{in}}$, which provides a way to feed the data to the nodes of the reservoir. These nodes are described with the annihilation operators $\hat b_j$. We assume a unidirectional cascade coupling from the source to the reservoir \cite{PhysRevLett.70.2273,PhysRevA.94.063825}. The evolution of the density matrix of the whole system, including both the reservoir and the source (input), is given by~\cite{ghosh2019quantum}
\begin{align}
       \begin{array}{*{20}{l}} {i\hbar \partial_t \rho } \hfill & = \hfill & {[ H_R,\rho ] + \frac{{i\gamma }}{2}\mathop {\sum}\limits_j {\cal{L}} ( \hat b_j) + \frac{{ip}}{2}\mathop {\sum}\limits_j {\cal{L}} ( \hat b_j^\dagger )} \hfill \\ {} \hfill & {} \hfill & { + i\mathop {\sum}\limits_{j,k} {f_{sk}} (t)W_{jk}^{{\mathrm{in}}}\left( {[ \hat a_{sk}\rho , \hat b_j^\dagger ] + [ \hat b_j,\rho  \hat a_{sk}^\dagger ]} \right)} \hfill \\ {} \hfill & {} \hfill & { + \,\frac{{i\eta }}{{2 }}\mathop  {\sum}\limits_{k} {f_k}  (t){\cal{L}}(\hat a_{sk})}, \end{array}\label{eq:djisdhf09f}
\end{align}
where $\rho$ denotes the density matrix of the entire system (source and reservoir), and we assume $\rho(t=0) = \rho_R \otimes \rho_s$, where $\rho_R$ is the initial density matrix of the reservoir, assumed to be in the vacuum state. The interaction with the environment is modeled using the Lindblad superoperator: ${\cal{L}}(\hat x) = 2\hat x\rho \hat x^\dagger - \hat x^\dagger \hat x\rho - \rho \hat x^\dagger \hat x
$ with $\hat{x} \in \{\hat{a}_{sk}, \hat{b}_j\}$. The function $f_k(t)$ describes the time-dependent coupling between the source mode, $\hat{a}_{sk}$, and the reservoir modes, $\hat{b}_j$. We assume a rectangular pulse function such that $f_{sk}(t) = 0$ except for the time interval $t^{(k)}_1 < t < t^{(k)}_2$ and $k$ stands for the number of the modes in the source. The last term in Eq. (\ref{eq:djisdhf09f}) accounts for the decay of source photons from the reservoir with a decay rate $\eta$.
We consider a reservoir consisting of a fully connected network of fermionic nodes coupled randomly according to the following Hamiltonian
\begin{align}
    H_{R}= \mathop {\sum}\limits_{ij} {J_{ij}} \left( { \hat b_i^\dagger  \hat b_j +  \hat b_j^\dagger  \hat b_i} \right),
\end{align}
where $\hat b_i\hat b_j^\dagger+\hat b_j^\dagger \hat b_i=\delta_{ij}$, the couplings $J_{ij}$ are randomly distributed within the interval $[-1, +1]$, and are normalized to the spectral radius, defined as the largest modulus of the eigenvalues of the Hamiltonian \cite{Cucchi_2022}. The reservoir may be incoherently pumped with a rate of $P$ to reach a steady state before coupling with the source modes. During the evolution, the average occupations of the reservoir nodes $\langle \hat b_j^\dagger \hat b_j\rangle$  are measured at different times selected to capture the complex dynamics of the reservoir. 
The information of the occupation numbers along with the corresponding $g^{(2)}$ form the feature vector $\mathbf{s}$ and the label (desired output), respectively. One can deduce that the output state of the reservoir $\mathbf{s}$ 
is a mapping from the input vector $\mathbf{u} := \rho_{si}$, where each element $\rho_{si}$ corresponds to the $i$-th sample from the dataset of initial source states. Then, learning (or modeling) implies that $\mathbf{g}^{(2)}(0) = \mathcal{W}(\mathbf{s})$, where $\mathbf{g}^{(2)}(0)$ is a vector with elements $g^{(2)}_i(0)$ (label of data) corresponding to the $i$-th input $\rho_{si}$, and $\mathcal{W}$ is a function that can be determined through training using classical machine learning regression
 \cite{colliot:hal-04225627}. That is, the feature data $\mathbf{s}$ is fed to a computer algorithm to train the model, which is later tested to check the accuracy of the model. At the core of the training process, we use an ensemble of decision trees. A decision tree is made up of a series of decision nodes that evaluate input currents for specific conditions \cite{machin_in_action}. Despite their simplicity, a single decision tree is prone to overfitting and tends to be unstable. To enhance performance while keeping the training simple, ensemble methods are used, that is, multiple decision trees are trained on randomly selected subsets of the data and of decision nodes, and their outputs are aggregated. This approach typically yields more accurate and stable predictions than those obtained from a single decision tree. To evaluate the performance of the modeling, we use the mean-square error (MSE) on the testing dataset, given by
\begin{align}
    \text{MSE}=\frac{\sum_i(D_i^{\text{test}}-D_i)^2}{\sum_i(D_i^{\text{test}}+D_i)^2}\,,
\end{align}
where $D_i$ is the ground truth value of $g^{(2)}(0)$ and $ D_i^\text{test}$ are $g^{(2)}(0)$ that the model estimates. 
\section{Results}\label{sec:res}
In this section, we consider different physical systems as the source, and for each case, we analyze the results of training and estimation of $g^{(2)}$. 
\subsection{Mixing thermal, coherent, and Fock states}
\begin{figure}
    \centering
\includegraphics[width=1.0\linewidth]{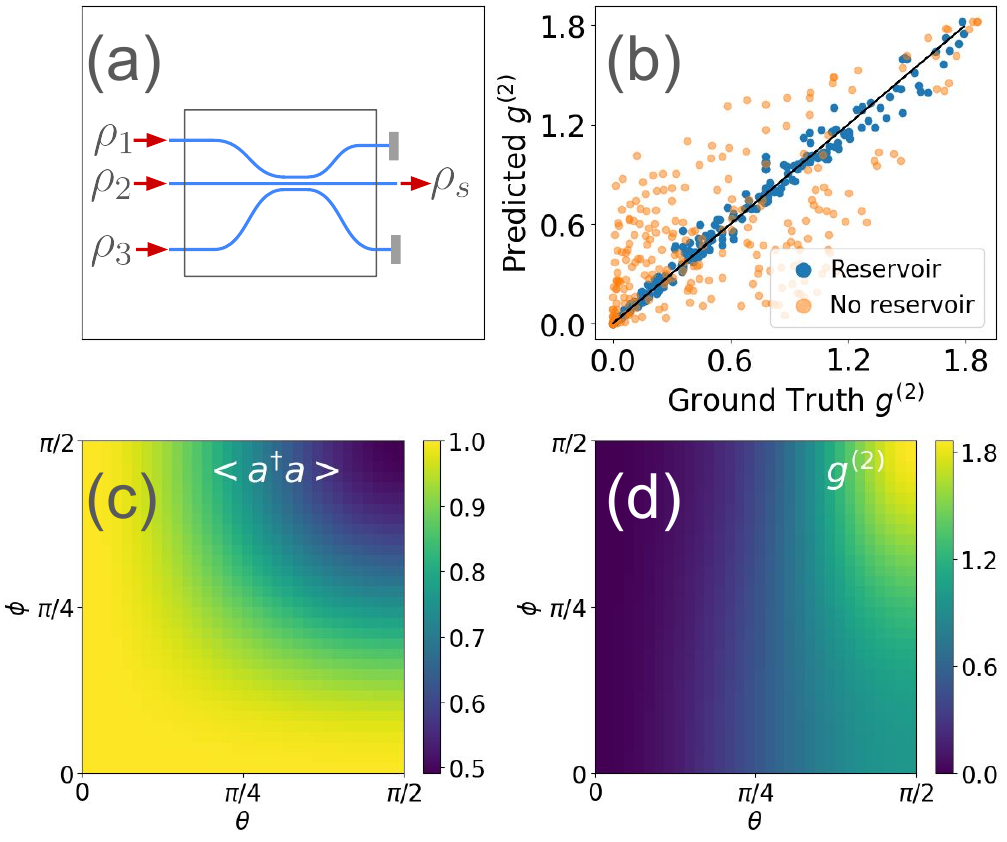}
    \caption{(a) Three-mode mixing in a setup consisting of three interacting guiding channels.  The density matrix $\rho_s$ corresponds to one of the output channels.
(b) Predicted vs. true values of $g^{(2)}(0)$ obtained by training a system that includes a quantum reservoir (blue dots) and a system processing density measurements directly (orange dots). The MSE of predictions is $0.0035$ and $0.154$, respectively.
(c) and (d) show the behavior of $\langle \hat a^\dagger \hat a \rangle$ and $g^{(2)}(0)$, respectively, as functions of the weighting parameters $\theta$ and $\phi$. 
Here we use $\bar n=0.5$ and $\alpha=1$.
}
    \label{fig:fih1mixing}
\end{figure}

A class of 
 states in which quantum information
tasks have been realized are CV states that can be derived from Fock states by coherently displacing and/or
squeezing them \cite{RevModPhys.77.513}. We consider a statistical mixture of CV quantum states: Fock, coherent, and thermal states. Such a combination can be realized in a three-mode interferometer, such as a tritter \cite{Spagnolo2013}, where three waveguides approach each other, interact, and then separate (schematic shown in Fig. \ref{fig:fih1mixing} (a).  The density matrix of our mixed source is $\rho_s=\sum_ip_i\rho_i$,
with $\rho_i$ being the density matrices of Fock, coherent, and thermal states. Parameters $p_i$ are weights (probabilities) fulfilling the condition $\text{Tr}[\rho_s]=1$. One way to quantify $p_i$ is through spherical coordinates: $p_1=\cos^2\theta$, $p_2=\sin^2\theta \cos^2\varphi$, and $p_3=\sin^2\theta \sin^2\varphi$ with $0\le\theta \le\pi$ and $0\le\varphi \le2\pi$.
Formal representations of density matrices in Fock basis are
\begin{align}
\rho_1=&\frac{(\hat a_s^\dagger)^n}{\sqrt{n!}}|0\rangle \langle0|\frac{(\hat a_s)^n}{\sqrt{n!}}= |n\rangle \langle n|\,,\\
    \rho_2= &e^{-\left|\alpha\right|^2} \sum_{n, m}^\infty \frac{\alpha^n \alpha^{*m}}{\sqrt{n! m!}} |n\rangle\langle m |\,,\\
    \rho_3=&\sum_{n=0}^\infty \frac{\bar{n}^n}{(\bar{n} + 1)^{n+1}} |n\rangle \langle n|\,,
\end{align}
where $\alpha$ is the complex amplitude  and $\bar{n}$ is the average photon number.
%
Given $\rho_s$ one can obtain   analytical expression for $g^{(2)}(0)$
\begin{align}
        g^{(2)}(0)=\frac{
    p_1n(n-1)
    +
    p_2|\alpha|^4
    +
    2p_3\bar{n}^2
    }
    {
    \big(
    p_1n
    +
    p_2|\alpha|^2
    +
    p_3\bar{n}
    \big)^2
    }\,.
    \end{align}

An example of the behavior of $g^{(2)}(0)$ is shown in Fig. \ref{fig:fih1mixing}(d). Due to the mixing of the three states, the value of $g^{(2)}(0)$ is in the range between well below 1 and up to values close to 2, corresponding to the regime of anti-bunching to super bunching, respectively. This range can be manipulated by adjusting the parameters $\bar{n}$ and $\alpha$. Using the quantum state corresponding to $\rho_s$ as the input to the reservoir and $g^{(2)}(0)$ as the target label for each training sample, we train the network to estimate the second-order correlation function.

Examples of predictions during the inference phase are shown in Fig.~\ref{fig:fih1mixing}(b).
Since in our scheme we collect the classical data (density in the nodes in the reservoir as a function of time), one may ask if we can estimate $g^{(2)}(0)$ directly from the density of the source state, without using the reservoir. To this end, we first generate a dataset consisting of  1024 samples, and then randomly split the dataset into a training set (with $75\%$ of the data) and test set (with $25\%$ of the data). Then, the classical data are fed directly to a nonlinear regression algorithm in the software, bypassing the quantum reservoir altogether. Our simulations show that this approach leads to much higher estimation errors, see scattered orange points in panel (b). This behavior can be inferred to some extent by scrutinizing panels (c) and (d), since there exist points exhibiting identical occupation numbers in the source that manifest distinct correlation values, thereby impeding the reliability of the estimation based on source density alone. On the other hand, with a reservoir consisting of two nodes only, the error is substantially reduced. In this case, the quantum state of the source is first fed into the reservoir, and the collected densities are used in the same nonlinear regression algorithm as before. An example of the prediction performance is shown in panel (b), where the testing results closely follow a straight line with a slope of one, indicating a strong agreement between predictions and ground truth values in the inference phase.     

\subsection{Emitter in a cavity}
Light modes can be confined in optical structures such as microcavities \cite{Vahala2003}, photonic crystals \cite{Taverne:18}, or plasmonic structures \cite{PhysRevLett.118.237401}. Embedding a quantum emitter whose excitonic transition resonates with a confined optical mode can lead to strong light–matter coupling, giving rise to a rich variety of quantum electrodynamical phenomena. Such interactions have enabled the development of advanced photonic devices, including optoelectronic devices \cite{Maryam2013} and single-photon sources \cite{Do2024}.  
\begin{figure*}
    \begin{center}
\includegraphics[width=1.0\linewidth]{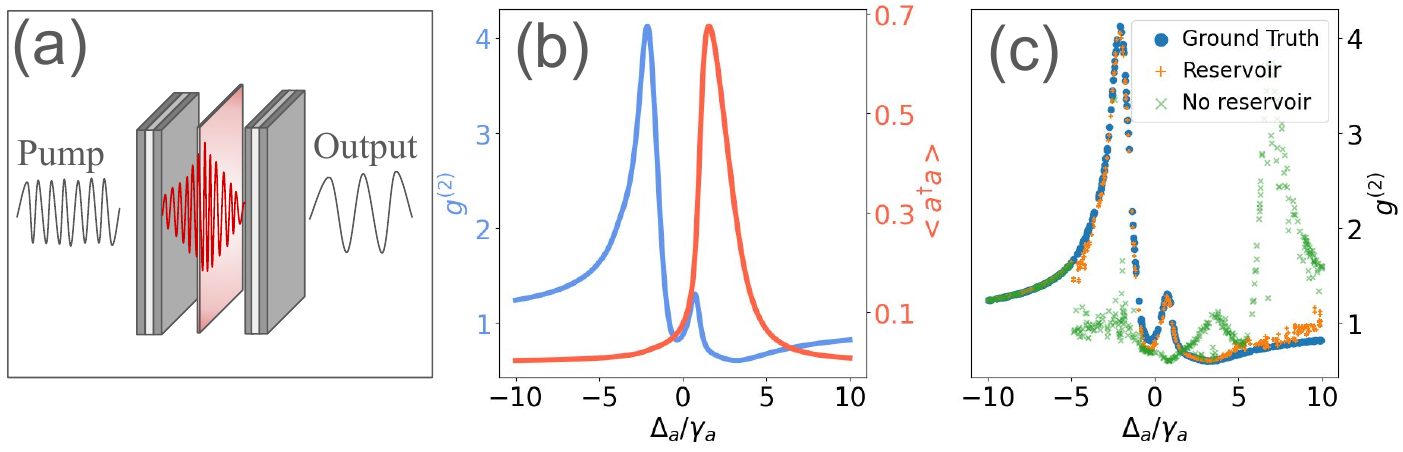}
    \caption{(a) The coupling between a bosonic emitter mode and a cavity photon mode can lead to interesting statistical behavior. (b) $g^{(2)}(0)$ and $n_a=\langle \hat a^\dagger \hat a\rangle$ of the source depicted in (a) as a function of detuning $\Delta_a$, modeled by Hamiltonian (\ref{eq:dhgf98wedghe}). Here, the parameters are scaled to the decay rate of the mode $\hat{a}$, i.~e.~$J/\gamma_a=1.6$, $U_b/\gamma_a=1$, $\gamma_b/\gamma_a=0.1$, $U_a=0$, $\Omega_a/\gamma_a=0.1$, $\Omega_b/\gamma_a=0.07$. (c) Predictions using a system with a two-node reservoir (shown by orange markers) and a system without a reservoir (shown by green markers). Here, MSE with reservoir is $0.037$ and without reservoir is $1.17$.}
    \label{fig:emitterincavity}
    \end{center}
\end{figure*}

To model such physics, we may consider the following Hamiltonian
\begin{align}\label{eq:dhgf98wedghe}
H_s = &  \Delta_a\, \hat{a}^\dagger \hat{a} 
    +  \Delta_b \, \hat{b}^\dagger \hat{b} 
    + J \, \left( \hat{a}^\dagger \hat{b} + \hat{a} \hat{b}^\dagger \right)\nonumber\\
    & + \frac{U_b}{2} \, \hat{b}^\dagger \hat{b}^\dagger \hat{b} \hat{b}+\frac{U_a}{2} \, \hat{a}^\dagger \hat{a}^\dagger \hat{a}\hat{a} \nonumber  \\
    & + \Omega_a  \hat{a} 
    + \Omega_b \hat{b} + h.c.\,,
\end{align}
where $\hat{a}$, $\hat{b}$ are bosonic operators, $\Delta_{a,b}/\hbar$ is the difference between frequency of the bosonic modes and the frequency of the pump,  $U_{a,b}$ are the coefficients of the Kerr nonlinearity and $J$ quantifies the coupling between the two modes. We assume that the system is coherently driven with rates $\Omega_{a,b}$. Depending on the nature of the modes, this model can describe different physical systems such as exciton-polaritons \cite{Khudaiberganov_2018}, coupled optical cavities \cite{Zhang:17}, or driven-dissipative Bose-Hubbard model \cite{Dahan2022}, to mention a few examples (see Fig.~\ref{fig:emitterincavity}(a) for a schematic of the structure). 
Assuming that mode $\hat{a}$ is accessible through photon detection, we are aiming to extract information about the second-order correlation function $g^{(2)}_a(0)$ associated with this mode, which encodes key statistical properties of the emitted light. We solve the master equation
    \begin{align}
        \partial_t\rho_s=&-\frac{i}{\hbar}[H_s,\rho_s]+\frac{\gamma_a}{2\hbar}(2\hat{a}\rho_s \hat{a}^\dagger-\hat{a}^\dagger \hat{a} \rho_s-\rho_s \hat{a}^\dagger \hat{a})\nonumber\\&+\frac{\gamma_b}{2\hbar}(2\hat{b}\rho_s \hat{b}^\dagger-\hat{b}^\dagger \hat{b} \rho_s-\rho_s \hat{b}^\dagger \hat{b})\,,
    \end{align}
where we introduce decay with rates $\gamma_{a,b}/\hbar$ in the Lindblad form. 

We now analyze an example of the training results for the system governed by the above Hamiltonian and master equation. As a first step, we study the behavior of the mean photon number $n_a$ and the second-order correlation function $g^{(2)}(0)$, both shown in panel (b) of Fig.~\ref{fig:emitterincavity}, plotted as a function of the detuning parameter $\Delta_a$. The average photon number $n_a$ remains low for large detunings—both positive and negative—indicating that the cavity is only weakly populated. However, a significant peak
in $n_a$ appears at positive detuning. This peak can be understood as a result of the resonant interaction between the pump mode and a hybridized normal mode arising from the coupling between the two photonic modes, that is,  when the detuning brings the bare mode into resonance with one of the hybrid modes, energy transfer from the driving field into the cavity is enhanced, leading to a substantial increase in the photon population.

Concurrently, the behavior of $g^{(2)}(0)$ across the detuning spectrum reveals distinct statistical regimes of the output light. For certain negative detuning, $g^{(2)}(0)$ exceeds 2, indicating a superbunching regime characteristic of strongly correlated photons. In contrast, as the detuning shifts to positive values, $g^{(2)}(0)$ can drop below 1, signaling antibunching behavior—a hallmark of nonclassical light. This transition between different photon statistics highlights the sensitivity of the system’s output to detuning~\cite{PDDrummond_1980}.
\begin{figure*}
    \centering
    \includegraphics[width=1.0\linewidth]{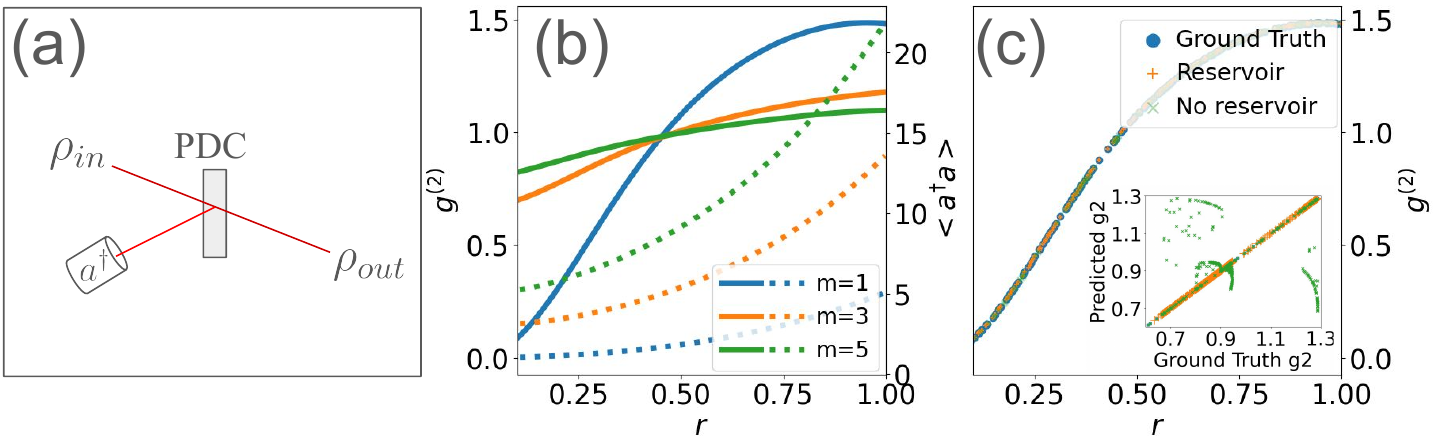}
    \caption{(a): Setup for generating photon-added squeezed states, where a photon is added to the squeezed state  $\rho_{in}$ in a parametric down conversion device (PDC).  (b) Example of the behavior of $g^{(2)}(0)$ (solid lines) and $\langle \hat a^\dagger \hat a \rangle$ (dashed lines) as a function of the squeezing parameter $r$. Both show a similar trend with respect to $r$. (c) Training a model with the data shown in panel (b) for $m=1$ (one photon-added) shows a good fit to the true value. Here, the MSE without the reservoir is $1.03 \times 10^{-6}$, and it changes very slightly when using the reservoir. The inset shows the case where the quantum reservoir improves the training while the classical regressor fails.}
    \label{fig:added-squeeze}
\end{figure*}

Based on these results, we select an appropriate parameter range for training $g^{(2)}(0)$. As discussed, its variation with $\Delta_a$ reveals rich and complex behavior, making it a challenging estimation task. As before, the quantum state corresponding to the $g^{(2)}(0)$ data is fed into the reservoir, which enhances the identification of meaningful patterns for training in the classical part. 
To proceed with the training, we generate a dataset consisting of  8000 samples, and randomly split the dataset into a training set (with $80\%$ of the data) and  test set (with $20\%$ of the data).
Before preprocessing the data in the reservoir, we first examine whether our classical machine learning estimator (ensemble decision tree) can infer a functional relationship between the photon number $n_a$ and the second-order correlation function $g^{(2)}_a$. 
The results of such classical regression (without reservoir) are shown in panel (c) with green points. While the regression performs well in a narrow range of very negative detuning, it shows significant errors within the majority of the detuning spectrum, leading to inaccurate predictions. This suggests that the input data from the source, without preprocessing in the reservoir, is not sufficient and highlights the necessity to process the quantum state through a quantum feature extraction mechanism, such as a reservoir.

Upon incorporating the reservoir, the model is able to successfully predict unseen values of $g^{(2)}(0)$, as demonstrated by the close agreement between predicted and true values shown in panel (c) using orange markers. The MSE is substantially reduced when the reservoir is included.

These findings confirm the essential role of the reservoir in preprocessing and transforming the quantum input into a higher dimension of the reservoir where learning becomes tractable by a classical algorithm, enabling accurate estimation of quantum statistical properties such as $g^{(2)}(0)$. 

\subsection{Photon-added squeezed states}
Another compelling class of non-classical states (compared to mixing CV states) that has obtained growing attention comprises photon-added and photon-subtracted states, which are generated through the successive application of photon creation or annihilation operators on a given state \cite{PhysRevA.43.492}. Starting with the single-mode squeezed vacuum
state, it was
shown that adding photons to the state can generate a non-classical state \cite{PhysRevA.89.043829}.
\begin{figure*}
    \centering
    \includegraphics[width=1.0\linewidth]{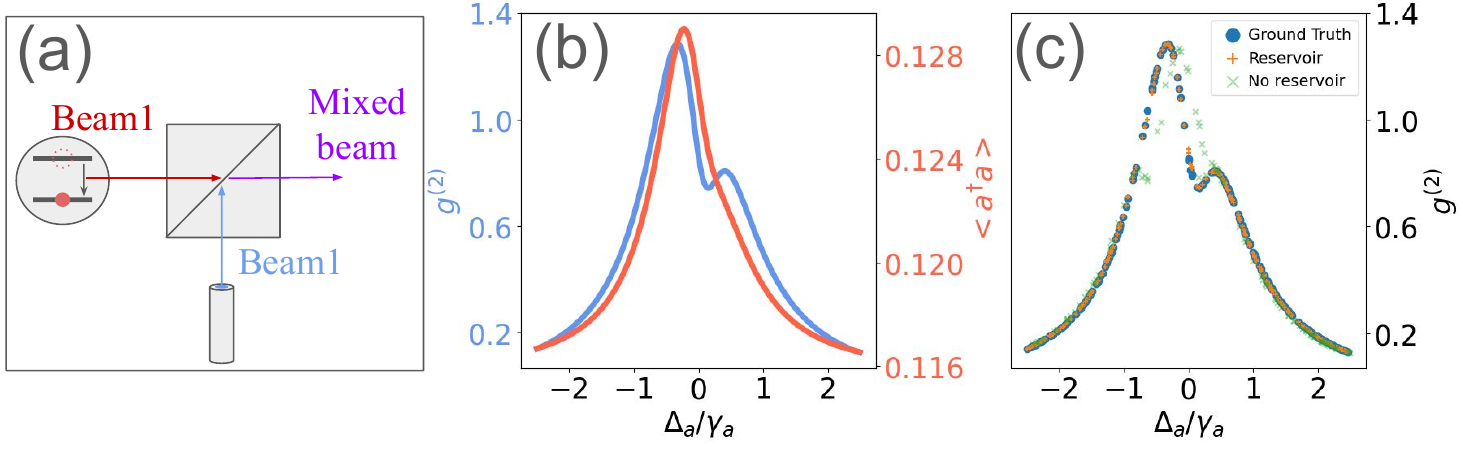}
    \caption{(a): In a beam splitter, two modes can be mixed to produce an output with interesting features. For example, a coherent state and the emission from a two-level system (e.g., fluorescence resonance) are mixed.
    (b) variation of $g^{(2)}(0)$ and $\langle \hat a^\dagger \hat a\rangle$ in the function of detuning $\Delta_a$. (c) Training a model within the range of parameters shown in panel (b) without reservoir (green crosses) and with a quantum reservoir (orange crosses). Here, MSE without reservoir is equal to $0.0089$ and with reservoir is equal to $7.47\times10^{-5}$.}
    \label{fig:2level-coh}
\end{figure*}

Here we consider the $m$ photon-added squeezed states as a representative example (see panel (a) in Fig. \ref{fig:added-squeeze}) and try to estimate its $g^{(2)}(0)$ using our proposal \cite{JClausen_1999,
JIANG2021127396, 
PhysRevA.93.052313}. The state is defined as
\begin{align}
    | \psi_{(r,m)} \rangle = N_{(r,m)}\hat a^{\dagger m} S(r) | 0 \rangle\,,
\end{align}
with $S(r)= \exp \left( \frac{r \hat a^2}{2} - \frac{r \hat a^{\dagger 2}}{2} \right)$. 

The normalization factor $N_{(r,m)}$ can be written as
\begin{equation}
N_{(r,m)}=1/\sqrt{m!\cosh^m{(r)}P_m(\xi)},    
\end{equation}
where $\xi=\cosh(r)$ and $P_m(\xi)$ is a Legendre polynomial. For this state, one can obtain $g_m^{(2)}(0)$ as
%
\begin{align}
g_m^{(2)}(0)=
\frac{(m+1)\xi\left[(m+2)\xi\frac{P_{m+2}(\xi)}{P_m(\xi)} -4\frac{P_{m+1}(\xi)}{P_m(\xi)}
 \right]+2}
 {
 \left[(m+1)\xi\frac{P_{m+1}(\xi)}{P_m(\xi)} - 1\right]^2
 }.    
\end{align}
Second order correlation function for single photon added squeezed state $g_1^{(2)}$ reads
\begin{equation}
g_1^{(2)}=\frac{3(3+2\tanh^2(r))}{\left[(\coth(r)-\tanh(r) )(3\cosh^2(r)-2)\right]^2}.    
\end{equation}

Examples of the behavior of the second-order correlation function $g^{(2)}(0)$ and the average photon number $n_a$ as functions of the squeezing parameter $r$ are presented in panel (b), Fig. \ref{fig:added-squeeze}, where both quantities increase monotonically with $r$. For the squeezed vacuum state itself ($m=0$), it can be shown that $g^{(2)}(0) = 3 + \frac{1}{\sinh^2{r}}$, which remains greater than one for all values of 
$r$, indicating super-Poissonian photon statistics and photon bunching.
However, the addition of photons to the squeezed vacuum state significantly modifies its statistical properties, especially when an odd number of photons is added. For instance, in the case of a single-photon-added squeezed vacuum state ($m=1$), $g^{(2)}(0)$ drops below one, entering the strong antibunching regime characteristic of nonclassical light.
As the squeezing parameter $r$ increases, 
$g^{(2)}(0)$ gradually increases, eventually transitioning back into the bunching regime. This behavior highlights the interplay between the nonclassical features introduced by photon addition and those inherent to the squeezed vacuum state itself.

Training data can be derived from the above equations. We generated 1000 samples dataset and sampled $20\%$ of it for testing purposes. An example of the training process for $m=1$ is shown in panel (c). As in the previous two cases, we initially train the model without a reservoir. Unlike the other cases, the inherent information from the source alone proves sufficient for learning the model. In this scenario, incorporating a reservoir appears unnecessary, as training with the reservoir leads to only a marginal decrease in the MSE. However, mixing data from different photon-added states (with varying values of \( m \)) and using them collectively to train a model makes it challenging to achieve reliable estimation based only on information from the source. An example of such an analysis is shown in the inset of panel (c) in Fig.~\ref{fig:added-squeeze}. In this case, the training is performed using quantum states with $ m = \{1, 3, 5\} $. When relying only on the source data, the training may result in a high MSE, as shown by the widely scattered green markers around the line of true values. Introducing the reservoir as a preprocessing framework dramatically improves the MSE, enabling the predictions to closely follow the line of true values. 
\subsection{Two-level system emission mixed with coherent state}
The last configuration we consider involves two beams entering a beam splitter (BS). The beam splitter can be described by a $2 \times 2$ unitary matrix, relating the input fields $(\hat{E}_1, \hat{E}_2)$ to the output fields $(\hat{E}_3, \hat{E}_4)$ as follows \cite{Gerry_Knight_2023}
\begin{align}
\begin{pmatrix}
\hat{E}_1 \\
\hat{E}_2
\end{pmatrix}
=
\begin{pmatrix}
iR & T \\
T & iR
\end{pmatrix}
\begin{pmatrix}
\hat{E}_3 \\
\hat{E}_4
\end{pmatrix}  \,. 
\end{align}
The elements $T^2$ and $R^2$ stand for the transmittance and reflectance of the BS, respectively, with the condition $T^2 + R^2 = 1$ ensuring energy conservation. Since the transformation between input and output fields must be unitary, the BS can equivalently be modeled by the unitary operator $U = e^{i\theta(\hat{E}_1^\dagger \hat{E}_2 + \hat{E}_1 \hat{E}_2^\dagger)}$, where the angle $\theta$ is defined by $R = \sin\theta$ and $T = \cos\theta$. The beam splitter is a fundamental component in many quantum technologies, including quantum information processing \cite{RevModPhys.83.33} and the generation of entangled states between input modes \cite{PhysRevLett.59.2044}. 
As we consider below, mixing a coherent light with the emission of a two-level system has interesting statistics \cite{PhysRevA.101.063824}.

In our setup (see panel (a) in Fig. \ref{fig:2level-coh}), one of the input beams to the BS is a coherent field, such as a laser beam, while the other originates from the emission of a two-level system (e.g., fluorescent light emitted by an atom). These two fields are mixed at the BS, and we use the output as an input to our device trained to estimate $g^{(2)}(0)$.

The dynamics of the coherent input field can be described by the following master equation
\begin{align}
\partial_t\rho_a=-\frac{i}{\hbar}[H_a,\rho]+\frac{\gamma_a}{2\hbar}(2a\rho_a \hat a^\dagger-\hat a^\dagger \hat a \rho-\rho \hat a^\dagger \hat a)\,,
\end{align}
with $H_a=\Delta_a \hat a^\dagger \hat a+\Omega_a(\hat a^\dagger+\hat a)$. The second input mode can be modeled as a two-level system using the following equation
    \begin{align}
        \partial_t\rho_\sigma=-\frac{i}{\hbar}[H_\sigma,\rho]+\frac{\gamma}{2\hbar}(2\sigma\rho_\sigma \sigma^\dagger-\sigma^\dagger \sigma \rho_\sigma-\rho_\sigma \sigma^\dagger \sigma).
    \end{align}
It can be realized by driving a two-level atom by a strong continuous-wave laser \cite{Scully_Zubairy_1997}, 
where the atom polarization operator is proportional to $\sigma$.  Here, $H_\sigma=\Delta_\sigma \sigma^\dagger \sigma+\Omega(\sigma+\sigma^\dagger) $. 
The two fields can be mixed in a beam splitter
\begin{align}
    o_1=&iR \sigma+T a\,,\\
    o_2=&T \sigma+iR a\,,
\end{align}
 Our goal is to consider the statistics of the output field $o_2$, that is, $g_{o_2}^{(2)}(0)=\frac{\langle o_2^\dagger o_2^\dagger o_2 o_2\rangle}{\langle o_2\dagger o_2\rangle^2}$ in order to extract meaningful information for training an estimator. In Fig.~\ref{fig:2level-coh}, panel (b), we present the variation of both $g_{o_2}^{(2)}$ and the mean photon number $n_{o_2}$ as functions of the detuning parameter $\Delta_a$. Choosing $\Delta_a$ as the control parameter does not restrict the possibility of exploring other parameters; it simply serves as an illustrative example. The results manifest a range of photon statistics, spanning from antibunching to bunching behavior in the output field. Notably, mixing at the BS can markedly transform the statistical properties of the fields. While a coherent light source exhibits $g^{(2)} = 1$, and a two-level emitter may produce $g^{(2)} \approx0$ \cite{PhysRevLett.125.170402}, the photon statistics of the output field can deviate considerably from either of these values. 

Next, we consider the learning processes, similar to the approach taken in the previous three cases. Panel (c) presents the results of the training.
When the system does not include a reservoir, we obtain a good fit between the predictions of the model and the actual data, except for the region of small positive detuning.  In this region, the occupation number in the source and $g^{(2)}$ show a very different trend, and this might be difficult for a classical estimator to make predictions using only occupation numbers of the source state. However, when the quantum reservoir is included in the system, the model accurately estimates $g^{(0)}$ in the entire range. This improvement is reflected in a reduced mean squared error and better training and testing scores.

\section{Discussion}\label{sec:disc}
A fundamental objective of machine learning is to enable reliable inference on data not seen in the training phase. This capability, commonly referred to as generalization, underpins the utility of machine learning models beyond their training domains \cite{Caro2022}. An important extension of this concept is to examine whether a model trained on quantum samples from one physical system can be effectively utilized to infer properties or make predictions about another, potentially dissimilar, system. Such cross-system applicability is particularly advantageous in contexts where data acquisition is limited, or when model training is substantially more demanding for one system compared to another. In these instances, the ability to transfer learned representations offers considerable benefits in terms of computational efficiency, resource allocation, and experimental feasibility.

Having the trained models of $g^{(2)}(0)$
  developed for various quantum optical systems discussed in the previous section, we now investigate their generalization capabilities. Specifically, we assess how well a model trained on samples from one system performs when evaluated on quantum states from a different source. To quantify this cross-system performance, we analyze the MSE obtained when applying a trained model to quantum states outside its training physical domain. Examples of this analysis are presented in Table \ref{table:diffcase}.  The diagonal numbers in Table \ref{table:diffcase} show the MSE obtained when training and testing are performed with the same source, that is, the physical system used during training is identical to that used for testing.  Notably, the model trained on the three-beam mixing configuration generalizes relatively well when applied to predict $g^{(2)}(0)$ for the emission of a two-level system mixed with a coherent state. However, its performance becomes significantly worse when evaluated on states from the emitter in the cavity, meaning a weaker capacity to be generalized in this domain, possibly highlighting fundamentally different quantum states. This is also the case when we examine MSE of other models that become worse with the emitter in the cavity source. 

\begin{table}[h!]
\caption{MSE analysis of 
cross-performance of a training model. Each row corresponds to the testing dataset and each column to the training model.}
\centering
\rowcolors{2}{white}{gray!10}
\resizebox{0.5\textwidth}{!}
{
\begin{tabular}{lcccc}
\toprule
        & \textbf{3B-Mix} & \textbf{Ph-Added} & \textbf{Em-in-Cav} & \textbf{Coh-2LS-Mix} \\
\midrule
\textbf{3B-Mix} & 0.0035 & 0.878070 & 1.557117 & 0.153847 \\
\textbf{Ph-Added} & 0.505315 & 0.000001 & 3.027573 & 0.124028 \\
\textbf{Em-in-Cav} & 0.990289 & 1.240838 & 0.037 & 1.641797 \\
\textbf{Coh-2LS-Mix} & 0.483793 & 0.370254 & 1.393258 & 0.000074 \\
\bottomrule
\end{tabular}
}
\label{table:diffcase}
\end{table}

\begin{figure}[h!]
    \centering
    \includegraphics[width=1\linewidth]{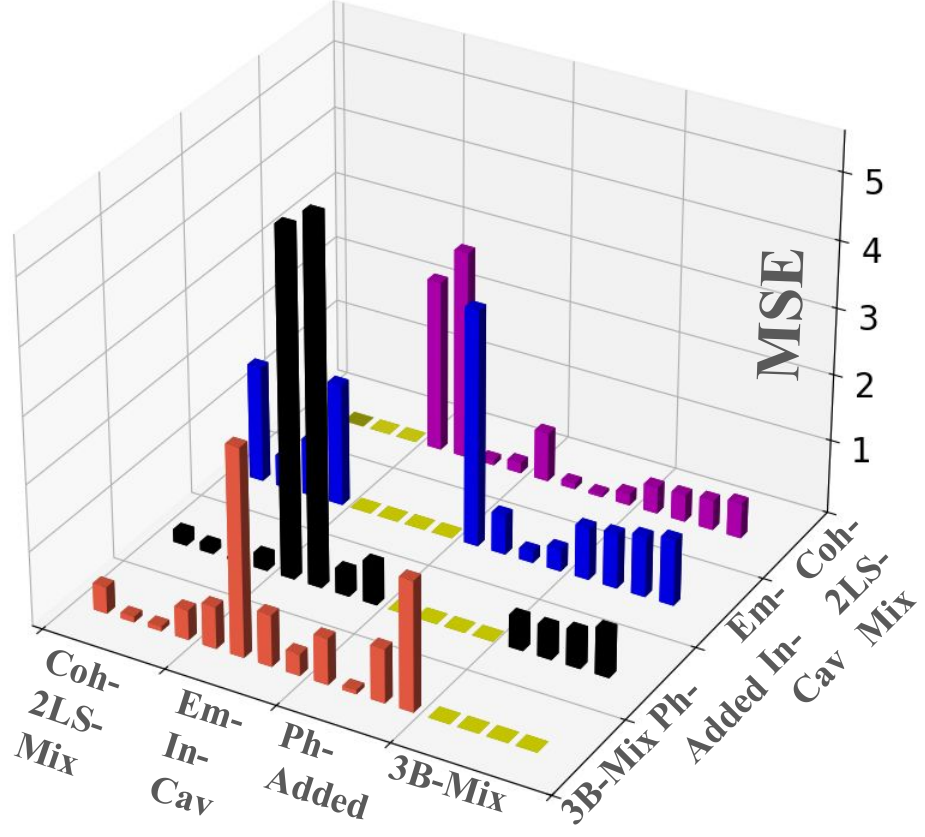}
    \caption{Partition-based MSE analysis. The true data for each physical model is divided into four segments, and the MSE is computed for each segment using a trained model.}
    \label{fig:chunk-anaalsijdois}
\end{figure}
\begin{figure}[t]
       \includegraphics[width=1.0\linewidth]{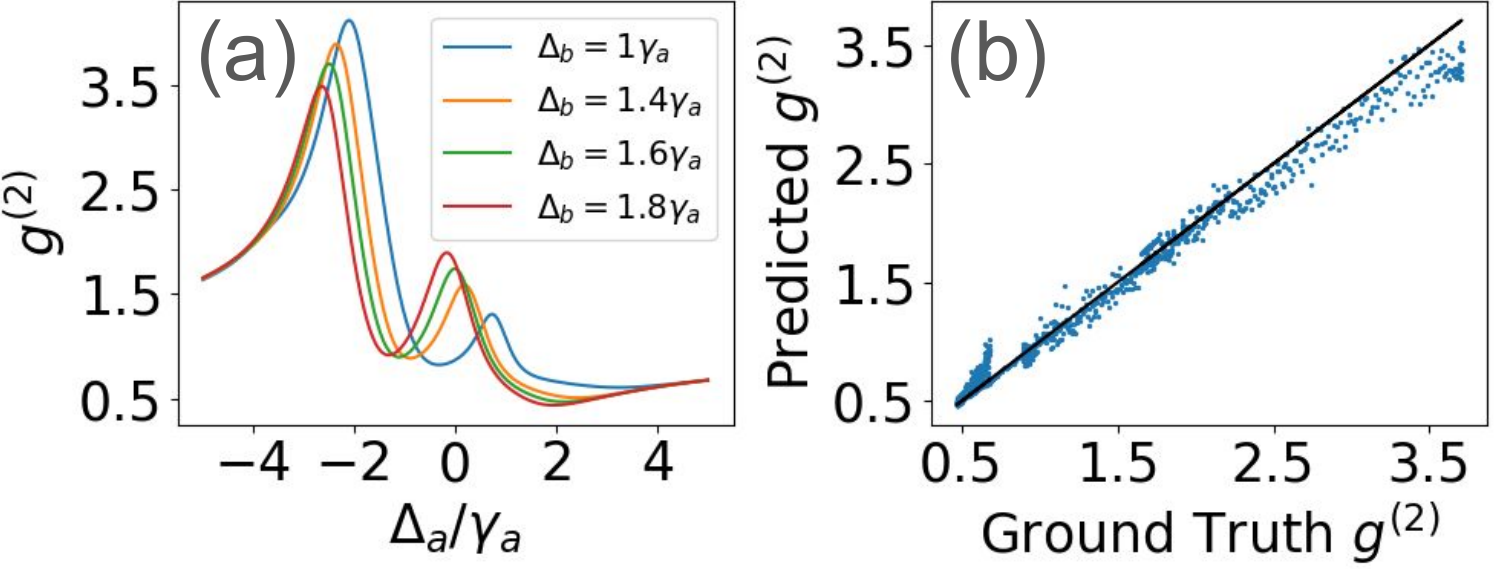}  \centering   \caption{Example of generalization learning in the case of an emitter in a cavity. (a) The second-order coherence function for different values of $\Delta_b$ as a function of $\Delta_a$. A model is trained using data for $\Delta_b = {1, 1.6, 1.8}\gamma_a$ and tested on unseen data with $\Delta_b = 1.4\gamma_a$. The testing result is shown in panel (c), with an MSE of $0.036$.}

    \label{fig:generaliajsdhuhd}
\end{figure}
To delve more into the generalization behavior of a trained model, we perform a partition-based MSE analysis. The true dataset is divided into four equal segments, and the MSE is evaluated for each segment using the pre-trained model. The results of this analysis are shown in Fig. \ref{fig:chunk-anaalsijdois}. When applying the model trained on the three-beam mixing (3B-mix) dataset to the true data of the one-photon-added (ph-added) dataset, we observe that the MSE decreases in certain segments, while remaining high in others—highlighting variability in model performance across data subsets. In the case of the emitter-in-cavity (em-in-cav) dataset, the MSE stays consistently high across all segments, showing persistent generalization challenges. Finally, for the coherent-state and two-level-system mixing (coh-2LS-mix) dataset, all segments exhibit relatively low MSE values compared to the two other cases, suggesting that the 3B-mix model generalizes more effectively to this physical configuration. This once again underscores that the em-in-cav configuration poses substantial difficulty for both training and generalization.

In Figure~\ref{fig:generaliajsdhuhd}, we illustrate an example that shows that some generalization capability 
for $g^{(2)}(0)$ 
can be achieved even for an emitter embedded in a cavity. In this case, instead of using very different systems, we consider similar systems but with different physical parameters in the training and testing phases.  We use input quantum states corresponding to various detuning values
$\Delta_b$. The model is trained on states for 
$\Delta_b = \{1.0, 1.6, 1.8\}\gamma_a$, and subsequently tested on unseen states for 
$\Delta_b=1.4\gamma_a$ during inference. The predicted outcomes, depicted in panel (b), demonstrate the model's ability to infer 
$g^{(2)}(0)$ for intermediate detuning values not included in the training set. This shows that our system can be used as a $g^{(2)}(0)$ measurement device for analyzing ``black box'' sources with unknown parameters, as long as the physical construction of the source is known, and similar to those used in the training phase.
%

\section{Conclusion}
In conclusion, our results demonstrated a proof-of-principle hybrid classical-quantum reservoir approach able to accurately estimate the zero-delay second-order coherence function $ g^{(2)}(0)$ across a variety of quantum light sources. While previous studies have explored hybrid classical-quantum schemes~\cite{PhysRevResearch.1.033063,10.3389/fphy.2022.1051941,Nokkala_2024,PhysRevA.107.062409}, our approach extends this framework by incorporating decision tree-based ensemble methods into the quantum reservoir. Moreover, we have explored the relevance of the out-of-distribution generalization ~\cite{Caro2023,PhysRevLett.133.050603}, where it may be applied to a certain regime of parameters. An exciting follow-up of our analysis is to employ the temporal processing ability of the quantum reservoir to estimate the time-delay second-order coherence function $ g^{(2)}(\tau)$. Such an extension would provide deeper insight into the underlying dynamics of quantum systems, including phenomena such as Rabi oscillations and relaxation behavior~\cite{Birnbaum2005,PhysRevA.111.023704}, as well as enable the characterization of nonclassical light sources exhibiting nonmonotonic photon correlations~\cite{PhysRevA.90.063824}.


%
\section*{Acknowledgment}
AR and MM acknowledge support from National Science Center, Poland (PL), Grant No.~2021/43/B/ST3/00752. DK acknowldges support from National Science Center, Poland (PL), Grant 2020/37/B/ST3/01657. AO acknowledge support from National Science Center, Poland (PL), Grant No.2024/52/C/ST3/00324. This work was financed by the European Union EIC Pathfinder Challenges project ``Quantum Optical Networks based on Exciton-polaritons'' (Q-ONE, Id: 101115575).
\bibliography{ref}
\end{document}